\documentclass[]{elsarticle}

\usepackage[usenames]{color}
\usepackage{bm}
\usepackage{amsmath}
\usepackage{amssymb}
\usepackage{epsfig}
\usepackage{graphics}

\def\be{\begin{equation}}
\def\en#1{\label{#1}\end{equation}}

\newcommand{\ran}{\rangle}
\newcommand{\eqb}{\begin{equation}}
\newcommand{\eqe}{\end{equation}}

\newcommand{\vare}{\varepsilon }

\begin{document}

\title{Nonlinear Zeno  dynamics   due to atomic interactions in Bose-Einstein condensate  }

\author{ V. G. Navarro}

\author{ V. S. Shchesnovich\corref{c}}
 \ead{valery@ufabc.edu.br; Tel.:+551149967960/Fax:+551149960090}\cortext[c]{Corresponding author}

\address{ Centro de Ci\^encias Naturais e Humanas, Universidade
Federal do ABC, Santo Andr\'e,  SP, 09210-170 Brazil }

\begin{abstract}

We show that nonlinear interactions induce both the  Zeno and anti-Zeno   effects in the  generalised Bose-Josephson model (with the on-site interactions and the second-order tunneling) describing   Bose-Einstein condensate in double-well trap subject to particle removal from one of the wells.  We find that the on-site   interactions   induce \textit{only} the Zeno effect, which appears at long evolution times, whereas   the   second-order  tunneling leads to a strong  decay of the atomic population at short evolution times,  reminiscent of the anti-Zeno effect, and  destroys   the nonlinear Zeno effect due to the on-site interactions at long times.

\end{abstract}

 \maketitle

\section{Introduction}

In  a seminal paper   Misra and Sudarshan \cite{MS}   introduced the quantum Zeno paradox by showing that an arbitrary evolution of quantum system comes to a  halt due to frequent measurements  (for a review, see Ref. \cite{FP}). The Zeno slowdown of the quantum evolution (in the decay of an  unstable  state)  was first observed in a two-level system \cite{IHBW}. Since then, understanding of the phenomenon has evolved significantly. It was rederived  as a purely dynamical effect (the Zeno effect) without the need for the von Neumann projection  postulate \cite{PN}. Moreover, besides the slowdown of   the  evolution of a system  (halted decay of an initial quantum state), i.e. the Zeno effect, enhancement of evolution, i.e. the anti-Zeno effect was also discovered \cite{KK,BR}. Generally speaking, both   effects are demonstration of an external control over  the quantum system, as it now  understood \cite{KK2}. There are   demonstrations of the Zeno and anti-Zeno dynamical effects  in open quantum systems,  where both effects  observed by varying the frequency of the observation (or the dissipative coupling rate), for instance, in cold sodium atoms \cite{FGR}, in spin-bath models \cite{SR}, and  in nanomechanical resonantor coupled to a point contact \cite{CTB}.  The Zeno effect   in a Bose-Einstein condensate (BEC) was first observed experimentally in Ref.~\cite{CPZeno}. Recently  the Zeno-like behavior in  a single    BEC  defect in an optical lattice  was recently considered theoretically \cite{ZenoDefTh} and observed experimentally \cite{ZenoDefExp}.

The above Zeno and anti-Zeno effects are due to an  \textit{external} influence on the system (i.e.  measurements or  a thermal bath, etc), however, the dynamical slowdown or enhancement of the system evolution can be also due to internal interactions in the system itself. Such effects  are also   counterintuitive, for instance, the inhibition of losses due to strong inelastic collisions in cold molecular gases \cite{InhLoss}.
It was also found that elastic collisions (i.e. ordinary nonlinear interactions) in a bosonic system, namely on-site elastic collisions between the BEC atoms  trapped in  a double-well potential,  lead to visible Zeno-type dynamics  \cite{VK}  which  was called   the nonlinear Zeno effect. In the latter system the Macroscopic Quantum Self Trapping (MQST) \cite{BES,MQST} appears simultaneously with the quantum Zeno dynamics, and, in principle,  could be   responsible for the nonlinear Zeno effect. It is thus of interest to study the respective range of parameters  where each of these two  effects appears, looking  for a domain of   the MQST, where   the nonlinear Zeno effect does not appear. Moreover, for the same reason, it would be  also interesting if one is able to demonstrate the anti-Zeno effect due to nonlinear interactions.

Therefore,  we set as the   main focus  of the present work   to study  the Zeno dynamics  due to two different types of nonlinear    interactions. Two different types of interactions are available already in the  most general (two-mode)    Bose-Josephson model of Ref.  \cite{AB}, describing a BEC in a double-well trap, where there are the on-site interactions in each mode and the second-order tunneling. One of the two modes is subject to an externally controlled  particle removal (loss), for instance,   due to application of an  electron   beam to one of the wells, similar as in Refs.  \cite{E1,E2,BKPO}.

We find that  the on-site interactions induce \textit{only} the Zeno  effect (at long evolution times), as compared to the non-interacting case. Moreover, we show that there are  parameter values for which    the MQST effect is expected, but  the nonlinear Zeno effect   does not appear, which was not   addressed  in our previous publication \cite{VK}.   On the other hand,   the second-order  tunneling   modifies the decay dynamics in a way   reminiscent of both    the  anti-Zeno effect at short evolution times and the Zeno effect at intermediate  evolution times.   Most importantly, the second-order  tunneling   can completely destroy the nonlinear  Zeno effect due to  the on-site interactions.

\section{ Bose-Josephson model   with an applied particle removal}

The most general two-mode  Hamiltonian,   describing BEC in an asymmetric double-well potential (i.e. the generalised Bose-Josephson model) was derived in Ref. \cite{AB}). It has  only two independent  parameters describing nonlinear interactions and   can be cast as follows  (equivalent to the Hamiltonian  $\hat{H}_2$ of Ref. \cite{AB})

\be
 H=-J(a_1^\dag a_2+a_2^\dag a_1)+\beta[(a_1^\dag a_2)^2+(a_2^\dag a_1)^2]+ V n_1
 +U(n_2^2 +n_1^2) +2(\beta-U)n_1n_2,
\en{E1}
where $a_j$ and $a_j^\dag$ ($j=1,2$) are the  boson operators of the two modes, $n_j = a^\dag_ja_j$, $J$ is the first-order    tunneling rate, $V$ is the zero-point energy bias,  $U$ is the local (on-site) nonlinear interaction  in each well,   and $\beta$ is second-order tunneling rate.   Note that $\beta-U$ is the strength of the nonlinear interactions across  the wells of the double well. We set  the  interaction between the atoms to be repulsive, i.e., $U>0 $, however,  the results   apply to the attractive case as well, due to a symmetry between the two cases (see, for details, Ref. \cite{STpaper}).    We also note that for a BEC in a double-well trap the on-site nonlinear interactions are always stronger than the second-order tunneling (see, for instance, Ref. \cite{AB}), i.e.,  $U > \beta$.

The  controlled removal of BEC atoms can be realized, for instance,   by using the   electron microscopy \cite{Ott_1,Ott_2} or by  a laser beam. In the former case, a narrow electron beam, ionizing the atoms, is directed to one of the minima of the potential.  In both cases, the applied removal is a continuous
measurement tool (the actual rate of the  condensate decay is directly observed) and can be described in the framework of the standard Markovian
approximation~\cite{BP}.   Introducing the  removal probability $p\equiv p(k_1,\Delta t)$, where $\Delta t$ is the  time interval and $k_m$ is a population of the $m$th
well, we can write the single atom removal event    as a quantum channel \cite{VK}:
\begin{equation}
|k_1,k_2\rangle|0\ran_R\to\sqrt{p}|k_1-1,k_2\rangle|1\ran_R+\sqrt{1-p
}|k_1,k_2\rangle|0\ran_R,
\label{QC}\end{equation}
where the atoms are removed from well 1,
$|k_1,k_2\ran=\frac{(a_1^\dag)^{k_1}(a_2^\dag)^{k_2}}{\sqrt{k_1!k_2!}}|0\rangle$ is
the ket-vector of the BEC state and $|j\ran_R$ describes the atom counter.
For the atom removal rate $\Gamma$ and a small $\Delta t$ we get  $p(k_1,\Delta t)\approx
\Gamma k_1 \Delta t$, where $\Delta t$ is   much less than the characteristic first-order  tunneling time, defined as
$t_\mathrm{QT} = {\hbar}/{J}$.

Introducing the reduced density matrix $\rho$ of  the BEC alone, the
quantum channel (\ref{QC}) can be described, for small times, by  the Kraus superoperator representation $\rho
\to M_0\rho{M}^\dag_0 + M_1\rho{M}^\dag_1$, where for a small $\Delta t$ we have
$M_0\approx 1-\Gamma n_1\Delta t/2$ and $M_1\approx \sqrt{\Gamma \Delta t\,}a_1$.
This leads to the master equation in the Lindblad form
\eqb
\frac{d{\rho}}{d t} = -\frac{i}{\hbar} [H,\rho]
+\Gamma\left\{a_1\rho{a_1^\dag}-\frac{n_1}{2}\rho-\rho\frac{n_1}{2}\right\}.
\label{EQ5}
\eqe
The  operator $\mathcal{D}(\cdot)=a(\cdot)a^\dag_1-\{n_1,(\cdot)\}/2$  has only
negative  eigenvalues $\lambda\in \{-N,-N+1/2,...,-1/2,0\}$, where $N$ is the total
number of atoms, i.e.    the only  stationary
state of  the operator $\mathcal{D}$ is  $\rho_0 = |0\rangle\langle0|$.

The  case of interacting bosons can be compared  to the non-interacting case, where the exact solution is readily available \cite{VK}. Indeed, for $U = \beta = 0$, setting also  $V=0$, for simplicity, and assuming that initially the condensate is in the ground state $|\psi\rangle =
\frac{(a^\dag_1+a^\dag_2)^{N_0}}{\sqrt{2^{N_0}N_0!}}|0\rangle$ one can derive the explicit solution \cite{VK}
\begin{eqnarray}
&&\langle N\rangle =e^{-\frac{\Gamma}{2}t}
\left[\frac{J^2}{(\hbar\Omega)^2}-\frac{\Gamma^2}{16\Omega^2}\cos(2\Omega
t)\right]N_0,
\label{EQ8}\\
&&\langle n_1-n_2\rangle = -e^{-\frac{\Gamma}{2}t} \frac{\Gamma}{4\Omega}
\sin(2\Omega t)N_0,
\label{n}\end{eqnarray}
where $\Omega=
\sqrt{\frac{J^2}{\hbar^2}-\frac{\Gamma^2}{16}}$.
Let us briefly recall how the     Zeno effect appears in the Bose-Josephson  model \cite{VK}. When $\Gamma<4J/\hbar$, Eqs. (\ref{EQ8}) and (\ref{n}) describe the decaying  Rabi
oscillations with the decay rate $\Gamma/2$. For $\Gamma\gg
4J/\hbar$, the dynamics is characterized by two different loss rates:  the initial
stage  with the rate $\Gamma/2$ and,  for times exceeding $1/\Gamma$, a
dramatically reduced dissipation rate   $\Gamma_\mathrm{QT} \approx
\frac{4J^2}{\hbar^2\Gamma}$.   The  inverse dependence of the actual loss rate  on a strong dissipation is the essence of the Zeno effect.  By introducing the   tunneling frequency   $\omega_R = 2J/\hbar$ one  obtains  the  following expression for the actual decay rate $\Gamma_{QT} = \omega_R^2/\Gamma$. We note that in the latter form our  decay rate  is  equivalent to that observed in  the continuous Zeno effect of Ref. \cite{CPZeno}.

Below we will use the following dimensionless control parameters: the normalized  applied rate $\gamma = \Gamma t_{QT} = \Gamma \hbar/J$, the nonlinear on-site interactions strength $\Lambda = UN_0/(2J)$, the   energy bias $\vare = V/J$, and the  second-order  tunneling rate   $\delta = \beta N_0/(2J)$,  where $N_0$ is the initial number of atoms.

\section{Mean-field approximation}
\label{sec2}
We find that in the limit of a large number of atoms $ N\gg 1$ one can   use the mean-field approximation obtained by decoupling the averages  as follows (see also Ref. \cite{VK}): $\langle n_j a_{j'}^\dag a_{j''}\rangle\approx\langle n_j\rangle\langle a_{j'}^\dag a_{j''}\rangle$. We have the following    mean-field variables  $z$, $\phi$, and $q$, which  correspond to the quantum averages:
\be
z=\frac{\langle n_1\rangle - \langle n_2\rangle}{\langle n_1\rangle + \langle n_2\rangle} , \quad
q=\frac{\langle n_1\rangle + \langle n_2\rangle}{N_0},\quad
e^{i\phi}=\frac{\langle a_1^\dag a_2\rangle}{\sqrt{\langle n_1\rangle \langle n_2\rangle}}.
\en{MFvar}
The mean-field equations read (with $\tau = t/t_{QT} = Jt/\hbar$)
\begin{equation}
\label{MF1}
\frac{dz}{d\tau}=-2\sqrt{1-z^2}\sin\phi-\frac{\gamma}{2}(1-z^2)+4\delta(1-z^2)q\sin(2\phi),
\end{equation}
\begin{equation}
\label{MF2}
\frac{d\phi}{d\tau}=\frac{2z}{\sqrt{1-z^2}}\cos\phi+\varepsilon+2(\Lambda - 2\delta) zq-4\delta zq\cos(2\phi),
\end{equation}
\begin{equation}
\label{MF3}
\frac{dq}{d\tau}=-\frac{\gamma}{2}q(1+z).
\end{equation}
The system of Eqs.~(\ref{MF1})-(\ref{MF3})  reduces to the mean-field equations of Ref. \cite{VK} when $\delta=0$. For $\gamma=0$ (and, hence, $q=1$) the system coincides with the mean-field system  of Ref. \cite{AB} and is Hamiltonian, i.e., $\dot{z} = -\frac{\partial \mathcal{H}}{\partial \phi}$ and $\dot{\phi} = \frac{\partial \mathcal{H}}{\partial z}$  with the Hamiltonian
\be
\mathcal{H} = (\Lambda-2\delta)z^2 + \varepsilon z-2\sqrt{1-z^2}\cos\phi + 2\delta(1-z^2)\cos(2\phi).
\en{Ham}

We have verified that  the  numerical solutions obtained by the Monte Carlo  method (a.k.a. the quantum jumps method) \cite{ME} are well approximated by   the mean-field  Eqs. (\ref{MF1})-(\ref{MF3}), where the Monte Carlo simulations were certified by comparison with the    exact solution (\ref{EQ8})  in the linear case.   We have found that there is a good agreement of the mean-field dynamics  with  the exact quantum dynamics already for   $N\gtrsim 10$. Therefore, below we use  the  mean-field Eqs. (\ref{MF1})-(\ref{MF3}) since they allow to significantly reduce the computation time (and  that they are an excellent approximation for   $N\gtrsim 100$).

\begin{figure}[ht]
\begin{center}
\epsfig{file=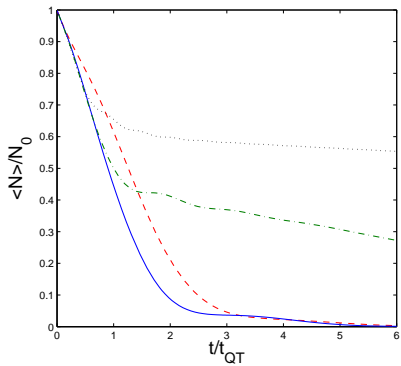,width=0.85\textwidth} \caption{ The average ratio of the number of particles in the system. Here $\gamma = 2$, and $\vare= \delta= 0$. We show the results of the numerical  simulations for $\Lambda=0$ (the solid line), $\Lambda = 5$ (the dashed  line), $\Lambda = 7$ (the dash-dotted line) $\Lambda = 10$  (the dotted line).     }
\label{F1}
\end{center}
\end{figure}

\section{Numerical simulations}
\label{sec3}

We first study the nonlinear Zeno effect due to the on-site interactions, first discovered in Ref.  \cite{VK},  by setting $\delta=0$ and $\vare = 0$ (for simplicity). The results are presented in Fig. \ref{F1}.  Observe that at short times
the interactions  leave  the behavior  of the   linear   (non-interacting) case (\ref{EQ8})  visibly  unmodified, whereas at long times the Zeno effect is clearly observed in comparison to the linear case for $\Lambda = 7$, for times $t\gtrsim 2 t_{QT}$, and for $\Lambda =10$, for times $t\gtrsim 1.5 t_{QT}$. At all values of the on-site interaction strength greater than the critical $\Lambda_{cr}= 1$ there is the MQST state in the closed system, i.e. for $\gamma=0$ \cite{MQST,STpaper}. On the other hand,  the  Zeno effect appears for larger $\Lambda>5$, since at $\Lambda = 5$ the average number of particles also reduces to zero (whereas if the MQST were responsible for the effect, it should have been finite). One thus concludes  that the MQST is not the mechanism responsible for the  remaining  finite fraction of the particles in the system  for $\Lambda = 7$ and $\Lambda  = 10$ in Fig. \ref{F1}. Hence,   the latter effect can be rightfully called  the nonlinear Zeno effect, which  confirmins  the main conclusion of Ref. \cite{VK}.
\begin{figure}[ht]
\begin{center}
\epsfig{file=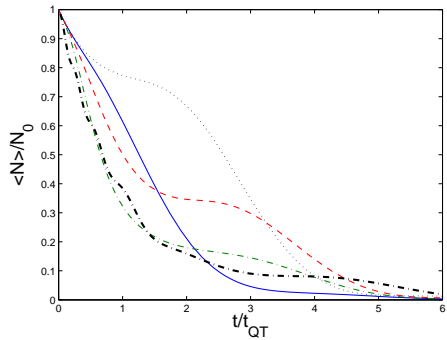,width=0.85\textwidth} \caption{ The average ratio of the number of particles in the system. Here $\gamma = 2$, and $\vare= \Lambda= 0$. We show the results of the numerical  simulations for $\delta=0$ (the solid line), $\delta = 1$ (the dotted line), $\delta = 2$ (the dashed line),  $\delta= 3$ (the dash-dotted line), and $\delta = 10$ (the tick dash-dotted line).     }
\label{F2}
\end{center}
\end{figure}

The influence of the second-order  tunneling on the actual decay rate  in the system is richer than that of the on-site interactions. To elucidate its influence, we consider first the second-order  tunneling alone by setting $\Lambda = 0$ (and $\vare = 0$ for simplicity).      The results are presented in Fig. \ref{F2}.  It is seen that the second-order  tunneling can result in both the Zeno and anti-Zeno effects, where the anti-Zeno effect appears for $\delta\ge 2$ at short evolution times, whereas the Zeno effect appears at intermediate  evolution times. Specifically,  for $\delta = 1$ we see no anti-Zeno effect, whereas a visible Zeno effect appears in the time window $ t_{QT} \lesssim t\lesssim 2t_{QT}$ for $\delta = 1$, in the time window $ 1.5t_{QT} \lesssim t\lesssim 3t_{QT}$ for $\delta = 2$  and $\delta = 3$, and in the time window $ 2t_{QT} \lesssim t\lesssim 5t_{QT}$ for $\delta = 10$. However, it is seen that the Zeno dynamics due to the second-order  tunneling is transient, since for long times the decay is comparable to that of the linear case.
\begin{figure}[ht]
\begin{center}
\epsfig{file=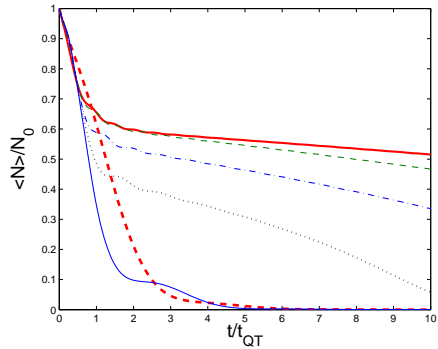,width=0.85\textwidth} \caption{ The average ratio of the number of particles in the system. Here $\gamma = 2$, and $\vare=   0$. We show the results of the numerical  simulations for $\Lambda=10$ and various values of $\delta$, with   $\delta=0$ (the thick solid line),    $\delta = 1$ (the thin dashed line), $\delta = 1.5$ (the dot-dashed line),  $\delta= 1.7$ (the  dotted line), and $\delta = 2$ (the  thin solid line) in comparison with $\Lambda=\delta=0$ (the thick dashed  line).     }
\label{F3}
\end{center}
\end{figure}

Finally, we study the effect of the combined action of the  on-site interactions and the second-order  tunneling. The results are presented in Fig. \ref{F3}. We see that the  second-order  tunneling destroys the nonlinear Zeno effect due to the on-site interactions. Moreover, there is a visible anti-Zeno dynamics due to the combined action of the second-order  tunneling and the on-site interactions (for instance, for $\Lambda=10$ and $\delta\ge 1.5$).
By examining  the mean-field Eqs.~(\ref{MF1})-(\ref{MF3}) one can see that the second-order  tunneling has two contributions: it induces an effective zero-point energy bias between the modes similar as the on-site interactions term, where the effective bias due to the  nonlinear interactions is $\vare_{NL} = 2(\Lambda - 2\delta)qz$. We note that for $\delta =0$ the latter effective energy bias explains the decay rate due to the  nonlinear Zeno effect derived in Ref. \cite{VK},
\begin{equation}
\Gamma_{NL}\approx\frac{4\Gamma}{ \gamma^2+4\vare^2_{NL}},
\end{equation}
which was shown to be an excellent approximation to the observed decay rate for   $t\gtrsim1/\Gamma$ provided that $ \gamma^2+4\vare^2_{NL}\gg 1$ \cite{VK}. Thus one would expect that for $\delta \ne 0$ the second-order  tunneling would  also contribute to the Zeno effect. However, the second-order  tunneling  has also the pair tunneling term which in the    mean-field system  (\ref{MF1})-(\ref{MF3})  is given by the last terms in Eqs. (\ref{MF2}) and (\ref{MF3}) (with the $\cos(2\phi)$ and $\sin(2\phi)$). We thus have found that this contribution dominates,   leading to the destruction of the Zeno effect due to the nonlinear interactions, resulting in a rapid decay of the atomic population comparable or faster than that of the linear (i.e. non-interacting)  case.

Here we note that  a sufficiently strong  second-order tunneling, due to contribution to  on-site interactions term, was previously found to modify the phase transition from the Mott isolator phase to the superfluid phase in the generalized Bose-Hubbard model. Indeed, it induces a new phase transition from a Mott insulator with one particle per site to a superfluid of spatially extended particle pairs living on top of the Mott background, instead of the usual transition to a superfluid of single particles/holes \cite{new1}.

\section{Conclusion}

We have studied the Zeno dynamics due to  two types of nonlinear interactions in the Bose-Josephson model, the on-site particle scattering and the second-order, i.e. pair, tunneling. We have found that   the on-site interactions induce only the nonlinear Zeno effect at long evolution times, which has    a larger threshold  value of the interaction   parameter than the Macroscopic Quantum Self-trapping. On the other hand, the second-order  tunneling is found to induce both the anti-Zeno effect at short evolution times and the Zeno effect at intermediate times. The latter effect is  transient, since at long evolution times the  atomic decay is similar to the linear case without nonlinear interactions.

Most importantly, we have found that the   second-order   tunneling  can lead  to  complete destruction of the Zeno effect due to the on-site interactions at long times leading to a strong decay of the atomic population, comparable or stronger that in the non-interacting case. This effect occurs even when there are    strong         on-site interactions   for  the  second-order  tunneling  rate comparable to that of the linear tunneling in the system. The  anti-Zeno effect  thus is a signature  of the second-order  tunneling in the Bose-Einstein condensate trapped in a double-well potential.

\section{Acknowledgementes}
The authors acknowledge the financial support by the CNPq and CAPES of Brazil.

\end{document}